# Planets, Planetary Nebulae, and Intermediate Luminosity Optical Transients (ILOTs)

**Noam Soker** [1,2,*]

[1] Department of Physics, Technion -- Israel Institute of Technology, Haifa 32000, Israel
[2] Guangdong Technion Israel Institute of Technology, Shantou 515069, Guangdong Province, China
* Correspondence: soker@physics.technion.ac.il

**Abstract:** I review some aspects related to the influence of planets on the evolution of stars before and beyond the main sequence. Some processes include the tidal destruction of a planet on to a very young main sequence star, on to a low mass main sequence star, and on to a brown dwarf. This process releases gravitational energy that might be observed as a faint intermediate luminosity optical transient (ILOT) event. I then summarize the view that some elliptical planetary nebulae are shaped by planets. When the planet interacts with a low mass upper asymptotic giant branch (AGB) star it both enhances the mass loss rate and shapes the wind to form an elliptical planetary nebula, mainly by spinning up the envelope and by exciting waves in the envelope. If no interaction with a companion, stellar or sub-stellar, takes place beyond the main sequence, the star is termed a *Jsolated star*, and its mass loss rates on the giant branches are likely to be much lower than what is traditionally assumed.

**Keywords:** Planetary systems; Planetary nebulae; stars: binaries; stars: AGB and post-AGB; stars: variables: general

## 1. Introduction

Planetary nebulae (PNe) can be shaped by stellar and sub-stellar companions (see Jones & Boffin 2017 for a recent review [1]). One open question that is with us for more than two decades (e.g., Soker 1996 [2]) is to what extent sub-stellar objects, and in particular planets, also shape PNe (see De Marco & Izzard 2017 for a recent review [3]). De Marco & Soker [4] took that about one quarter of all stars in the initial mass range $1-8M_\odot$ do form PNe, and estimate that about 20% of all PNe were shaped via planets and brown dwarfs. This amounts to about 5% of all $1-8M_\odot$ stars. In light of the general interest in the manner by which planets can influence stellar evolution (e.g., [5-10]), I discuss some issues related to star-planet interaction. The paper is based on a talk I gave at the Asymmetrical Planetary Nebulae (APN) VII meeting (Hong Kong, December 2017), and all figures are from my presentation at the meeting.

## 2. Engulfment of Planets by Asymptotic Giant Branch (AGB) Stars

For a planet to influence the envelope of an AGB star the envelope cannot be too massive. This implies a low mass star. However, for traditionally used mass loss rates low mass stars reach very large radii already on their red giant branch (RGB; Figure 1), and are likely to swallow close planets before they even reach the AGB. The way to have planets interact with their host star on the AGB is if the mass loss rate on the giant branches is lower than what traditional values are. In that case, the stellar core on the AGB and consequently the stellar radius are larger than in standard theoretical calculations, and the star is much more likely to swallow a planet on its upper AGB. Sabach & Soker [12,13] assume that *Jsolated stars*, i.e., those that are not spun-up in their post-main sequence evolution, lose mass at a rate that is less than about one third of the traditional one (for justification see [12]). They then show that in that case low mass AGB stars reach much larger radii and are much more likely to swallow planets than in standard theoretical calculations. The assumption of lower mass loss rates of Jsolated stars needs further examination by future observations and theoretical studies.

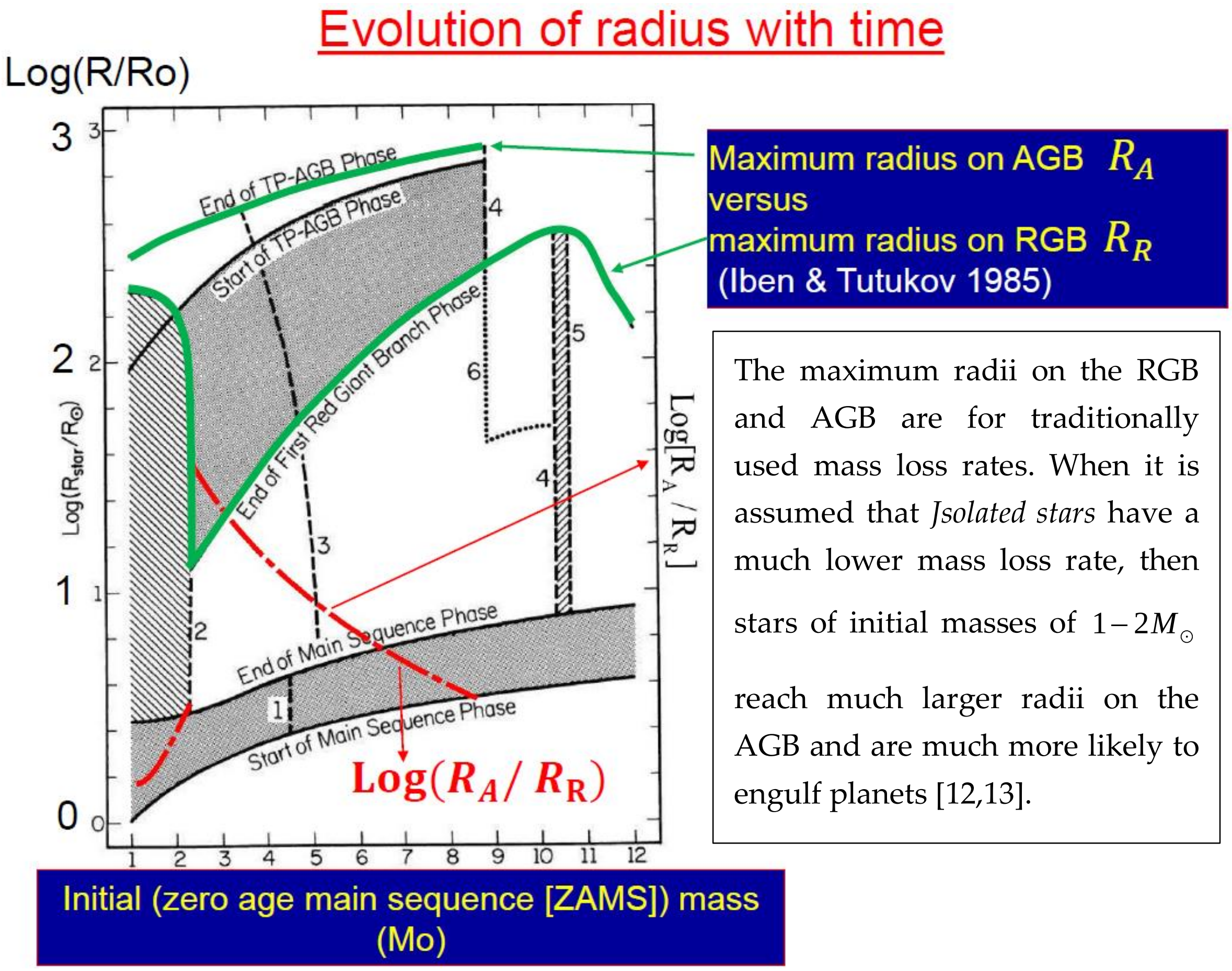

**Figure 1.** Maximum radii stars reach on their RGB and AGB as function of their initial mass for traditional mass loss rates (from [11]).

## 3. The Fate of a Planet: Tidal Destruction versus Engulfment

The fate of the planet as it comes close to the envelope of a star depends on the density ratio. If the density of the planet is larger than that of the star it dives in to the envelope as one entity and starts a common envelope evolution [14,15]. It will later be destroyed near the core, by either tidal forces or evaporation. Light planets are evaporated before they reach the core. Planets that are more massive can reach closer to the core and then suffer tidal destruction with part of their material accreted on to the core (marked "destruction on core" in Figure 3). If the density of the planet is lower than that of the star, it is tidally destroyed and forms an accretion belt (or a disk) around the star. In Figure 2 I present a schematic evolution of the planet and the stellar radii and densities, and mark which of the two outcomes takes place. I give more details of the outcomes in Figure 3.

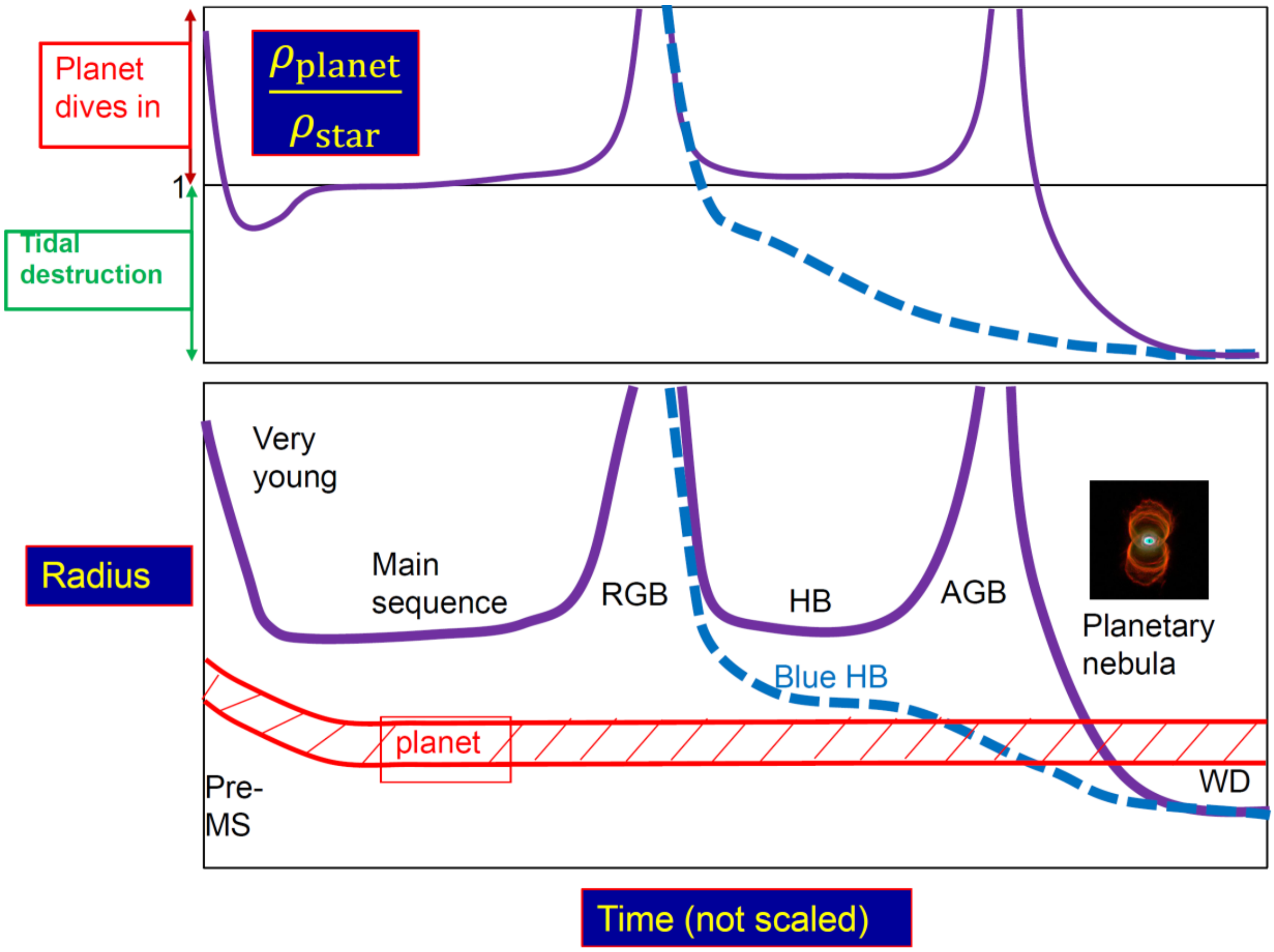

**Figure 2.** A schematic evolution of the radii and densities of planets and stars from the pre-main sequence phase to the WD phase. Upper panel: the ratio of the planet density to the stellar density. If the ratio is above 1 the planet dives-in to the envelope as one entity. If the density ratio is below 1 the planet is tidally destroyed and forms and accretion belt/disk around the star. Lower panel: The planet and stellar radii as function of time.

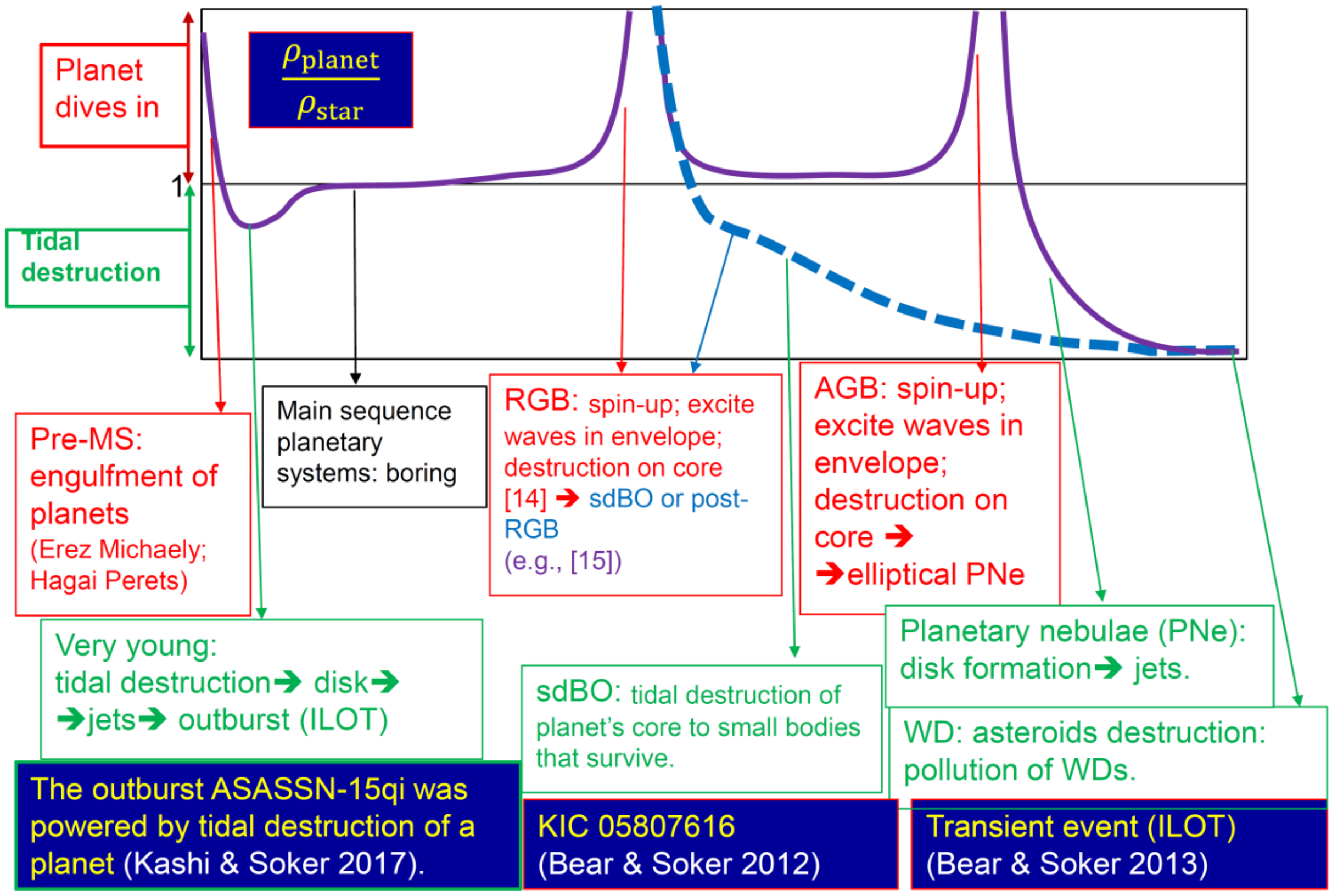

**Figure 3.** Outcomes of planet-star interaction.

## 4. Intermediate Luminosity Optical Transients (ILOTs) with planets

ILOTs are outbursts of a star or a binary system with a peak luminosity mostly between those of novae and supernovae (other names for these events are red novae, luminous red novae, and intermediate luminous red transients). Several studies have proposed that the interaction of a planet with a star can account for a minority of ILOTs. Retter & Marom (2003 [16]) proposed that V838 Mon was a result of planets entering a common envelope with a star. Bear et al. (2011 [17]) proposed that the destruction of a planet by a brown dwarf or a low mass main sequence star can result in an ILOT event. Kashi & Soker (2017 [18]) proposed that the outburst of the young stellar object ASASSN-15qi was an ILOT event where a sub-Jupiter young planet was tidally destroyed by a young main-sequence star. Because the system was young, the density of the planet was smaller than that of the star (Figure 3), and the planet was tidally destroyed. This, they suggested, resulted in the formation of an accretion disc and a gravitationally powered ILOT. The mass of the planet was too small to inflate a giant envelope, and hence the ILOT was hot, rather than red. As well, its energy was below those of classical novae.

Bear et al. (2013 [19]) discussed the possibility of observing the transient event that might result from the tidal destruction process of an asteroid near a WD. However, this event is much weaker than typical ILOTs.

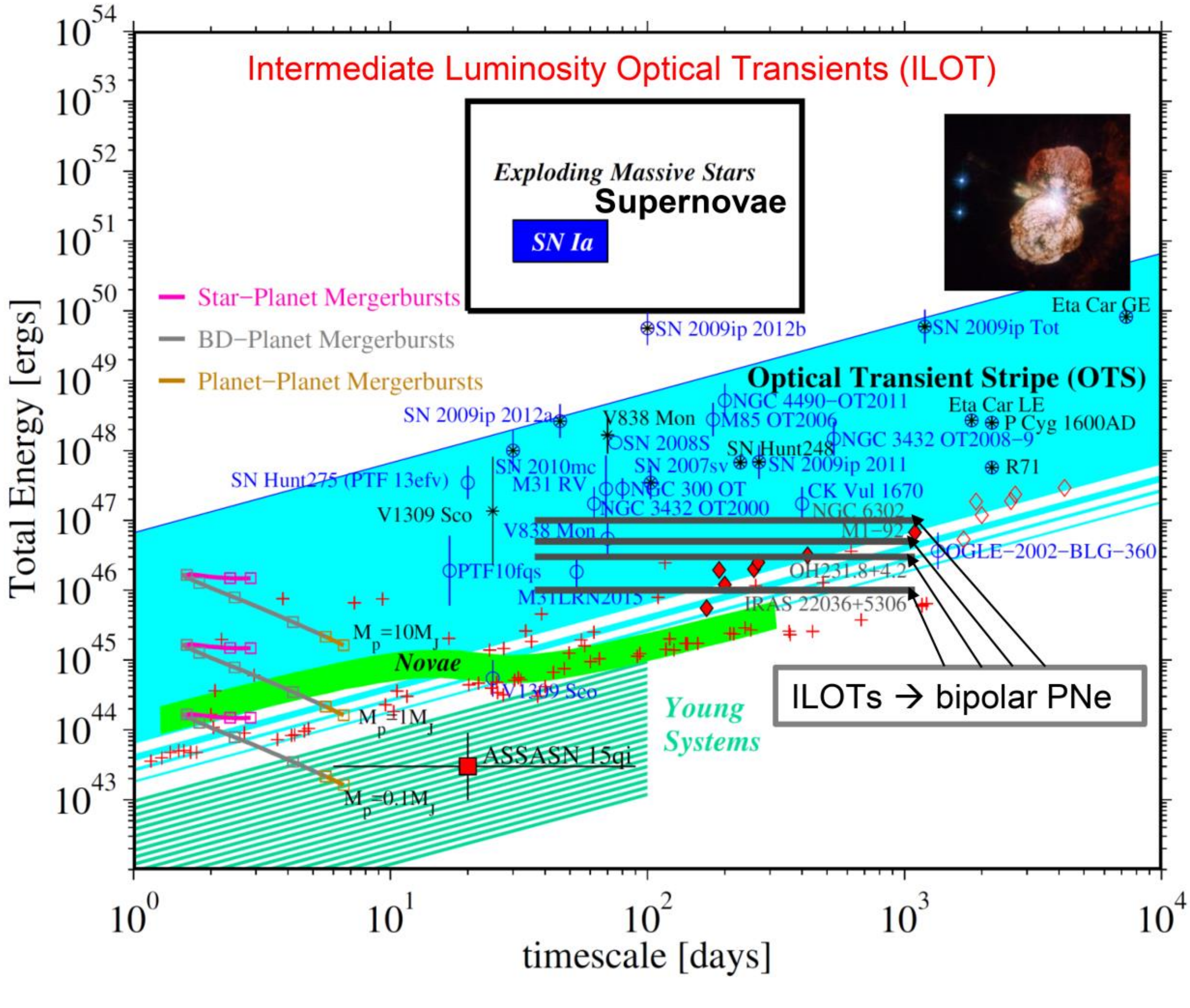

**Figure 4.** Observed transient events on the energy time diagram. Blue empty circles represent the total (radiated plus kinetic) energy of the observed transients as a function of the duration of their eruptions, i.e., usually the time for the visible luminosity to decrease by 3 magnitudes. The Optical Transient Stripe is populated by ILOT events that we [18] suggest are powered by gravitational energy of complete merger events or vigorous mass transfer events. For details of this figure see http://phsites.technion.ac.il/soker/ilot-club/

## 5. Intermediate Luminosity Optical Transients (ILOTs)

In figure 4 I present the energy-time diagram of ILOTs that Amit Kashi and I have been developing in the last several years (see http://phsites.technion.ac.il/soker/ilot-club/ for an updated diagram).

We suggest ([18] and references therein) that these ILOTs are powered by gravitational energy in one of several types of processes. (1) The secondary star is completely destroyed and part of its mass is accreted onto the primary star, e.g., as a planet destruction onto a brown dwarf. (2) The secondary star enters the envelope of a companion but stays intact and forms a common envelope, e.g., as Retter & Marom (2003 [16]) suggested. (3) The secondary star accretes mass while outside the envelope of the primary star, e.g., as in our model for the Great Eruption of Eta Carinae or our suggested scenario for some PNe [18].

With Amit Kashi [20] we suggest that the binary progenitors of some bipolar PNe experienced ILOT events that shaped the PN. The several months' long to several years' long outbursts were powered by mass transfer from an AGB star on to a main sequence companion that orbits outside the AGB envelope. Jets launched by an accretion disk around the main sequence companion shaped the bipolar lobes. Four such bipolar PNe are marked on Figure 4. They are marked with a long horizontal line because we know more or less the kinetic energy of the nebulae, which is about the ILOT energy, but we cannot tell how long the mass ejection phase lasted.

## 6. Planet-Shaped Planetary Nebulae

When a planet spirals-in inside the loosely bound envelope of an upper AGB star it can excite waves in the envelope and spin-up the envelope, both of which can cause asymmetrical mass loss. Finally, when the planet is destroyed near the core it might lead to further asymmetrical mass loss from inside the envelope, e.g., jets that might be launched by the core. For example, a Jupiter-like planet might form a disk that launches jets with about few percent of the mass of the planet. This amounts to a jets' mass of about $\mathrm{few} \times 10^{-5} M_{\odot}$. This mass is sufficient to form two opposite bullets (`ansae') along the symmetry axis of the nebula.

I started the paper with the discussion of the general interaction of planets with evolving stars, I then moved to discuss the formation of ILOTs with planet companions, and in section 5 I mentioned some bipolar PNe that can be shaped by ILOT events with a stellar companion to the AGB progenitor of the PN. I now end the discussion of planet-shaped PNe by listing the evolutionary channels and the resulting PN types. For that I use a table from De Marco & Soker (2011 [4]), which I present here as Figure 5.

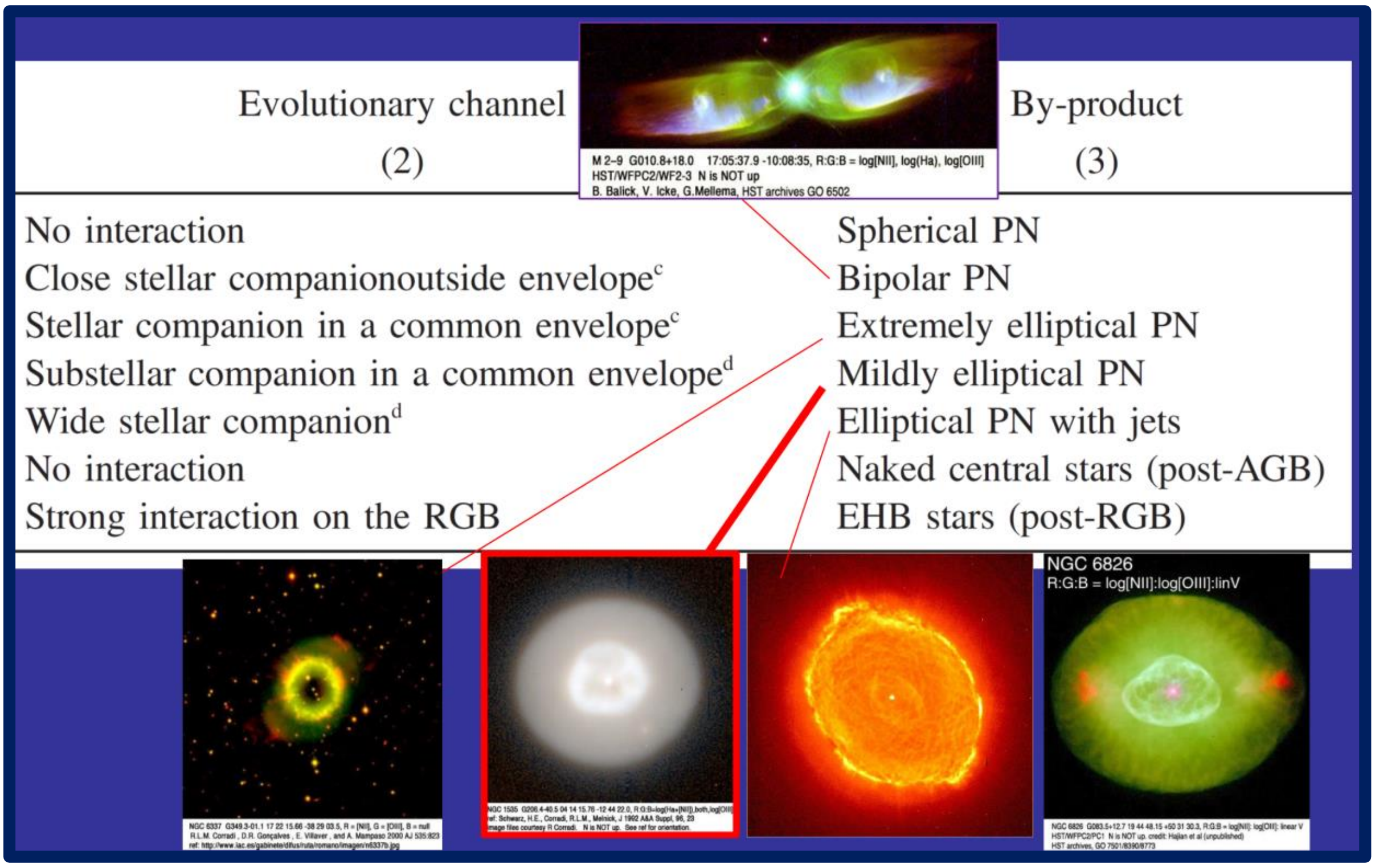

**Figure 5.** The evolutionary channels and the PN types that result from them (based on De Marco & Soker 2011 [4]).

## 7. Discussion

I discussed some aspects of the influence of planets on late stellar evolution, and its relation to some aspects of stellar binary interaction. The main claim of this presentation is that planets can influence the evolution of low mass stars, in, e.g., enhancing the mass loss rate of RGB and AGB stars (that without any companion are *Jsolated stars* that have low mass loss rates, lower than what is usually assumed), in forming some ILOTs, and in shaping PNe. There are other aspects I did not get into, such as the strength of the tidal interaction between the planet and the star, some aspects of which can be found in reviews from earlier related meetings (e.g., [21, 22]).

I thank the referees for useful comments.